\documentclass[preprint]{aastex}
\usepackage{color,natbib}
\newcommand{\Msun}{\mbox{$M_{\sun}$}}

\newcommand{\Lsun}{\mbox{$L_{\sun}$}}

\newcommand{\Mjup}{\mbox{$M_{\rm Jup}$}}
\newcommand{\Rjup}{\mbox{$R_{\rm Jup}$}}

\newcommand{\perone}{\mbox{$^{-1}$}}

\newcommand{\etal}{et al.}
\newcommand{\eg}{e.g.}
\newcommand{\ie}{i.e.}
\newcommand{\cf}{cf.}

\newcommand{\kms}{\hbox{km~s$^{-1}$}}
\newcommand{\chisq}{\mbox{$\chi^2$}}

\newcommand{\meth}{{\hbox{CH$_4$}}}   % CH4
\newcommand{\Ks}{\mbox{$K_S$}}
\newcommand{\degs}{\mbox{$^{\circ}$}}
\newcommand{\Lbol}{\mbox{$L_{\rm bol}$}}

\newcommand{\Teff}{\mbox{$T_{\rm eff}$}}
\newcommand{\logg}{\mbox{$\log(g)$}}

\newcommand{\bPic}{\hbox{$\beta$}~Pic}

\newcommand{\ps}{\protect\hbox{Pan-STARRS1}}
\newcommand\TPS{$3\pi$~Survey}
\newcommand{\gps}{\ensuremath{g_{\rm P1}}}
\newcommand{\rps}{\ensuremath{r_{\rm P1}}}
\newcommand{\ips}{\ensuremath{i_{\rm P1}}}
\newcommand{\zps}{\ensuremath{z_{\rm P1}}}
\newcommand{\yps}{\ensuremath{y_{\rm P1}}}

\newcommand{\WISE}{{\sl WISE}}

\newcommand{\psobject}{PSO~J318.5$-$22}
\newcommand{\psobjectfull}{PSO~J318.5338$-$22.8603}

\slugcomment{ApJ Letters, in press}
\shorttitle{An Extremely Red Young L Dwarf from PS1}
\shortauthors{Liu et al.}

\begin{document}

\title{The Extremely Red, Young L~Dwarf \psobjectfull: A Free-Floating
  Planetary-Mass Analog to Directly Imaged Young Gas-Giant Planets}


\author{Michael C. Liu\altaffilmark{1,2},
Eugene A. Magnier\altaffilmark{1},
Niall R. Deacon\altaffilmark{3},
Katelyn N. Allers\altaffilmark{4},
Trent J. Dupuy\altaffilmark{5},
Michael C. Kotson\altaffilmark{1},
Kimberly M. Aller\altaffilmark{1},
%and PS1 Builders
W. S. Burgett\altaffilmark{1}, 
K. C. Chambers\altaffilmark{1}, 
P. W. Draper,\altaffilmark{6}   % Durham
K. W. Hodapp\altaffilmark{1}, 
R. Jedicke\altaffilmark{1}, 
N. Kaiser\altaffilmark{1}, 
R.-P. Kudritzki,\altaffilmark{1}
N. Metcalfe,\altaffilmark{6}    % Durham
J. S. Morgan\altaffilmark{1}, 
P. A. Price\altaffilmark{7},    % Princeton
J. L. Tonry,\altaffilmark{1} 
R. J. Wainscoat\altaffilmark{1}
}

\altaffiltext{1}{Institute for Astronomy, University of Hawaii, 2680
  Woodlawn Drive, Honolulu HI 96822}
\altaffiltext{2}{Visiting Astronomer at the Infrared Telescope Facility,
  which is operated by the University of Hawaii under Cooperative
  Agreement no. NNX-08AE38A with the National Aeronautics and Space
  Administration, Science Mission Directorate, Planetary Astronomy
  Program.}
\altaffiltext{3}{Max Planck Institute for Astronomy, Konigstuhl
17, D-69117 Heidelberg, Germany}
\altaffiltext{4}{Department of Physics and Astronomy, Bucknell
  University, Lewisburg, PA 17837}
\altaffiltext{5}{Hubble Fellow. Harvard-Smithsonian Center for
  Astrophysics, 60 Garden Street, Cambridge, MA 02138}
%----------------------------------------
\altaffiltext{6}{Department of Physics, Durham University, South Road,
  Durham DH1 3LE, UK}
\altaffiltext{7}{Department of Astrophysical Sciences, Princeton
  University, Princeton, NJ 08544, USA} 
% %

\begin{abstract} 
  We have discovered using \ps\ an extremely red late-L dwarf, which has
  \mbox{$(J-K)_{MKO} = 2.84$} and \mbox{$(J-K)_{2MASS} = 2.78$}, making
  it the reddest known field dwarf and second only to 2MASS~J1207$-$39b
  among substellar companions.
  Near-IR spectroscopy shows a spectral type of L7$\pm$1 and reveals a
  triangular $H$-band continuum and weak alkali (\ion{K}{1} and
  \ion{Na}{1}) lines, hallmarks of low surface gravity.
  Near-IR astrometry from the Hawaii Infrared Parallax Program gives a
  distance of 24.6$\pm$1.4~pc and indicates a much fainter $J$-band
  absolute magnitude than field L~dwarfs.
  The position and kinematics of \psobject\ point to membership in the
  \bPic\ moving group. Evolutionary models give a temperature of
  1160$^{+30}_{-40}$~K and a mass of 6.5$^{+1.3}_{-1.0}$~\Mjup, making
  \psobject\ one of the lowest mass free-floating objects in the solar
  neighborhood.
  This object adds to the growing list of low-gravity field L~dwarfs and
  is the first to be strongly deficient in methane relative to its
  estimated temperature.
  Comparing their spectra suggests that young L~dwarfs with similar ages
  and temperatures can have different spectral signatures of youth.
  For the two objects with well constrained ages (\psobject\ and
  2MASS~J0355+11), we find their temperatures are $\approx$400~K cooler
  than field objects of similar spectral type but their
  luminosities are comparable, \ie, these young L~dwarfs are very red
  and unusually cool but not ``underluminous.''
  Altogether, \psobject\ is the first free-floating object with the
    colors, magnitudes, spectrum, luminosity, and mass that overlap the
    young dusty planets around HR~8799 and 2MASS~J1207$-$39.
\end{abstract}


%----------------------------------------------------------------------%
\section{Introduction}

A major surprise arising from direct detection of gas-giant planets
around young stars is that the spectral properties of these objects
differ from those of field L~and T~dwarfs
\citep[e.g.][]{2005A&A...438L..25C, marois08-hr8799bcd,
  bowler10-hr8799b, 2010A&A...517A..76P, 2011ApJ...733...65B,
  2013ApJ...774...55B}. These young planets have redder near-IR colors,
fainter near-IR absolute magnitudes, and peculiar spectra compared to
their field analogs. Over the last several years, this development
  has fostered closer scrutiny of the long-standing paradigm that a
  simple physical sequence connects the lowest mass stars to brown
  dwarfs to gas-giant planets.

We now know that most field brown dwarfs are not good analogs to
  young exoplanets. In contrast to field T~dwarfs of similar
  temperature, the young planets around HR~8799 and 2MASS~J1207$-$39
  have no methane absorption and very red colors. These properties are
  thought to arise from extreme atmospheric conditions tied to their
  young ($\approx$10--30~Myr) ages and low gravities, \eg, enhanced
  vertical mixing, non-equilibrium chemistry, and unusual clouds
  \citep[e.g.][]{2011ApJ...735L..39B, madhu11-8799,
    2012ApJ...754..135M}. This is corroborated by recent studies of the
  youngest ($\sim$10--100~Myr) field brown dwarfs, which find that
  low-gravity L~dwarfs also show very red colors and spectral
  peculiarities \citep[e.g.][]{2003ApJ...596..561M,
    2008ApJ...689.1295K}. Complicating this interpretation, however, is
  the existence of very red L~dwarfs that do not show spectral
  signatures of youth \citep[e.g.][]{2010ApJS..190..100K,
    allers13-young-spectra}. Thus, similarities between the colors and
  spectra of field brown dwarfs and young planets can have ambiguous
  interpretations.

There are two prominent shortfalls in our observational knowledge.
  (1)~There are only a handful of very red young L~dwarfs currently
  known (see compilation in \citealp{2012AJ....144...94G}), and only one
  of them has a parallax measurement \citep{2013AJ....145....2F,
    2013AN....334...85L}. (2)~Most L~dwarfs do not have as red near-IR
  colors as young exoplanets, and {\em none} are as faint in their
  near-IR absolute magnitudes. Therefore, the utility of young field
  objects as exoplanet analogs may be limited, since the existing
  samples of these two types of objects are small and do not really
  overlap.

We have found an extraordinary young L~dwarf that will help shed
  light on these topics. It is the reddest field object found to date
  and the first to have absolute magnitudes comparable to directly
  imaged young dusty exoplanets.

%----------------------------------------------------------------------%
\section{Observations \label{sec:observations}}

We have been undertaking a search for T~dwarfs using the \ps\ (PS1)
\TPS\ \citep{deacon11-ps1-tdwarfs, 2011ApJ...740L..32L}. 
We select objects using \ps\ 
and 2MASS
based on their colors and proper motions
and then obtain near-infrared photometry for further screening.
We observed \psobjectfull\ (hereinafter \psobject) using WFCAM
on the UK Infrared Telescope (UKIRT) on the 2010 September~15~UT.
Conditions were photometric with 0.9--1.0\arcsec\ seeing.
We found that the $(J-K)_{MKO}$ color for \psobject\ was
$2.74\pm0.04$~mag (Table~1), significantly redder than any previously
known field dwarf.

We obtained $R\approx$100 near-IR (0.8--2.5~\micron) spectra on
2011~July~21~UT from NASA's Infrared Telescope Facility.
(By coincidence, we observed \psobject\ immediately before obtaining the
spectrum of the similarly red L~dwarf WISE~J0047+68 published in
\citealp{2012AJ....144...94G}.) Conditions were photometric with
1\arcsec\ seeing. We used the near-IR spectrograph SpeX
\citep{1998SPIE.3354..468R} in prism mode with the 0.8\arcsec\ slit.
The total on-source integration time was 16~min.
All spectra were reduced using version~3.4 of the SpeXtool software
\citep{2003PASP..115..389V, 2004PASP..116..362C}.
We also used the final spectrum to synthesize near-IR colors for
\psobject\ (Table 1).

To assess the gravity of \psobject, we obtained $R\approx1700$ near-IR
spectra using the GNIRS spectrograph \citep{2006SPIE.6269E.138E} on the
Gemini-North 8.1-m Telescope.
Cross-dispersed spectra were obtained using the 
0.3\arcsec\ slit and the 32~lines~mm\perone\ grating
on the nights of 2013 June~26, June~30, and July~01 UT.
The total integration time was 5400~s. We reduced the data using a
version of SpeXtool modified for GNIRS cross-dispersed data.
We combined the telluric-corrected spectra from the three nights using a
robust weighted mean to produce the final 0.95--2.5~\micron\ spectrum.

We conducted astrometric monitoring of \psobject\ with the facility
near-IR camera WIRCam
at the Canada-France-Hawaii Telescope, obtaining 9~epochs over 2.0~years
starting on 2011~July~26~UT.
Our methods are described in \citet{2012arXiv1201.2465D}.
Using 116 reference stars in the field of \psobject, the resulting
median astrometric precision per epoch was 4.0\,mas, and the best-fit
proper motion and parallax solution had $\chi^2 = 13.2$ with 13 degrees
of freedom.
We applied a relative-to-absolute parallax correction of
$0.74\pm0.13$\,mas derived from the Besan\c{c}on model of the Galaxy
\citep{2003A&A...409..523R}.  Table~1 gives our astrometry results.
We did not find any objects co-moving with \psobject\ in our
10$\farcm$4$\times$10$\farcm$4 field of view within a range between
0.6\,mag fainter and 3.3\,mag brighter at $J$~band.

%----------------------------------------------------------------------%
\section{Results \label{sec:results}}

\subsection{Spectrophotometric Properties \label{sec:spectra}}

\begin{figure}
\vskip -0.3in
\hskip -0.4in
\hbox{
  \vbox{\hsize=4in
    \includegraphics[width=3in,angle=0]{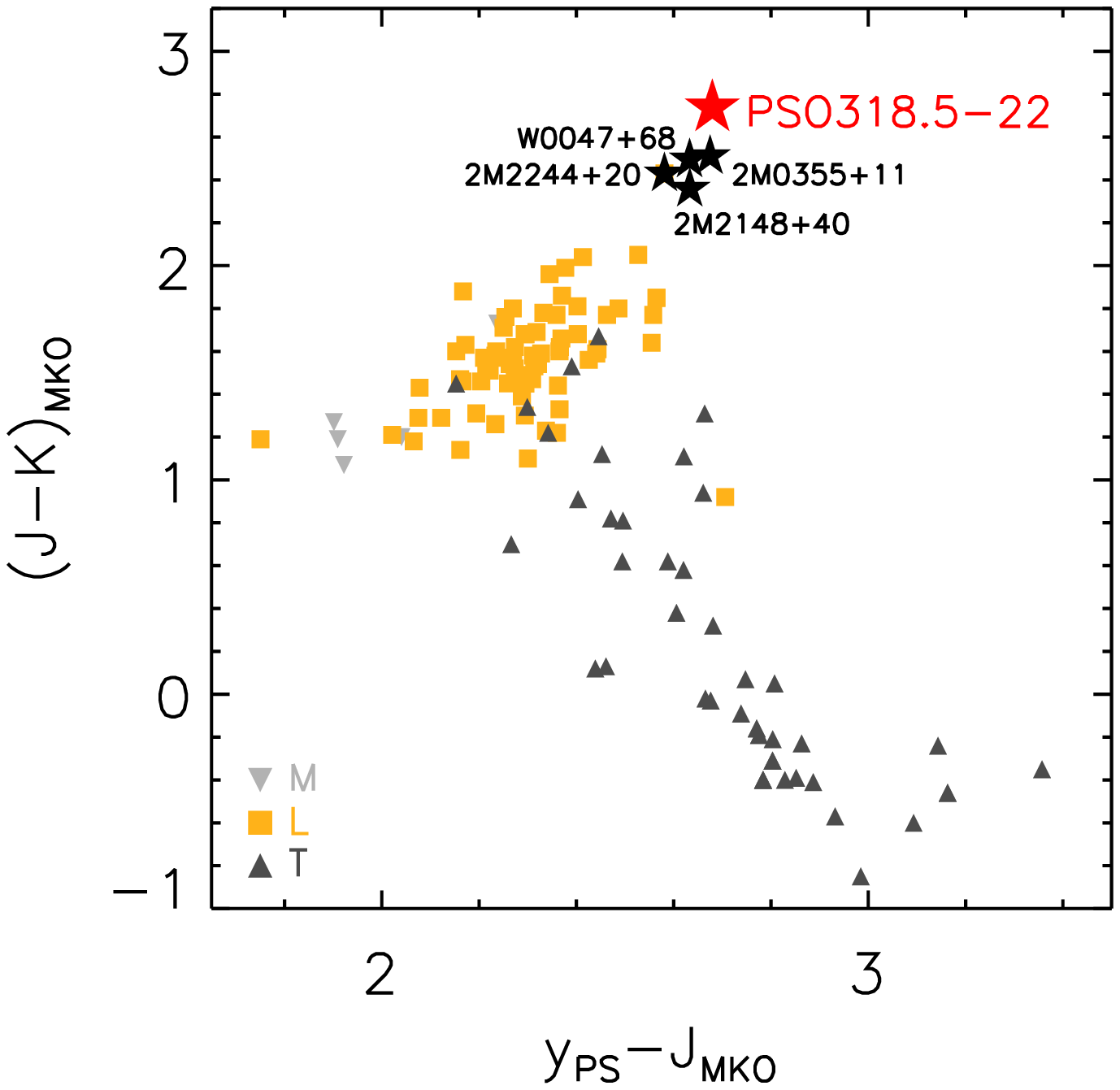}
    \vskip -0.2in
    %\hskip -0.17in
    %\includegraphics[width=3in,angle=0]{deacon-JHHK.ps}
    \includegraphics[width=3in,angle=0]{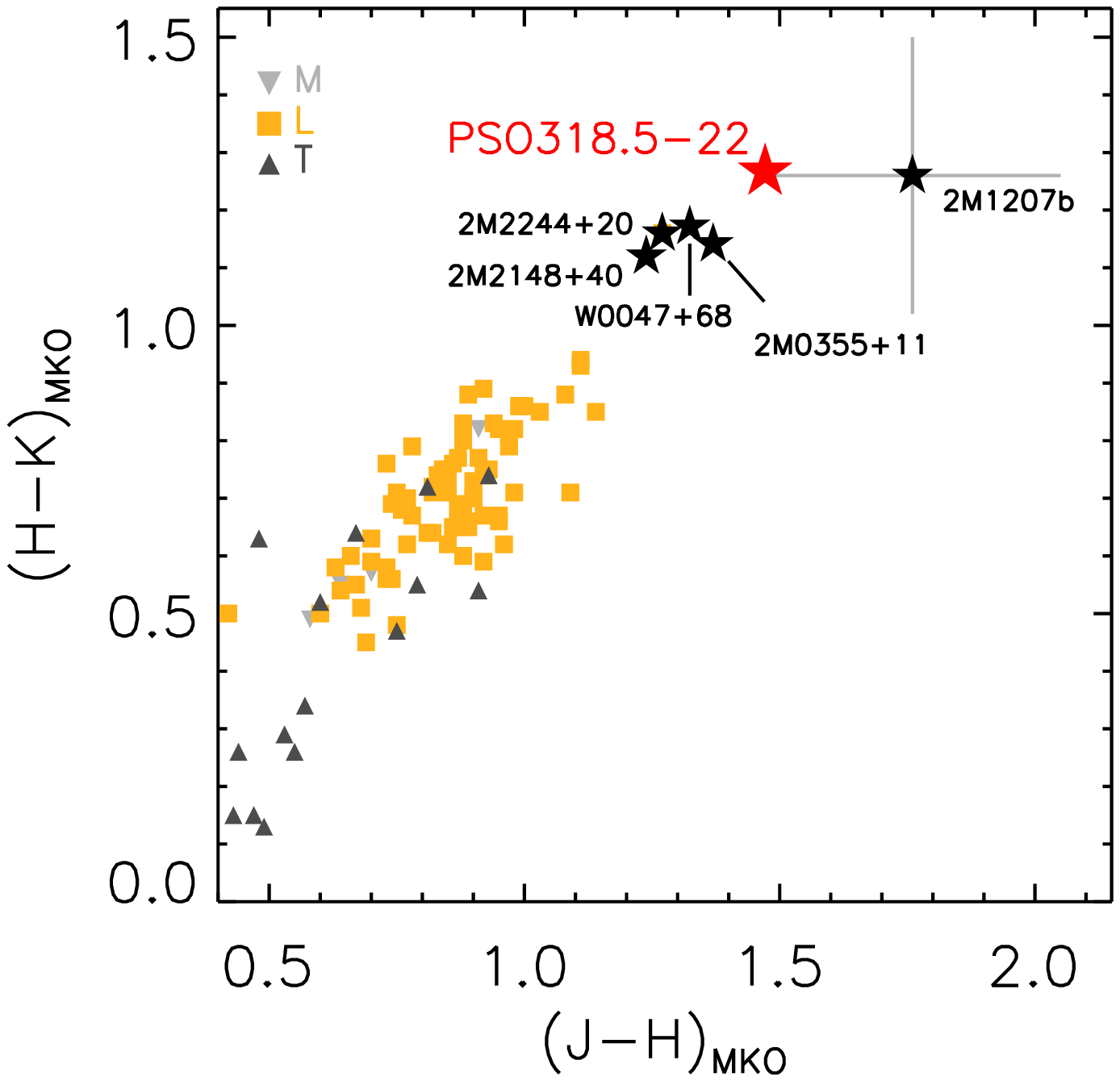}
  }
  \hskip -3.2in
  \raise -0.38in
  \vbox{
    \includegraphics[height=9.3in,angle=90]{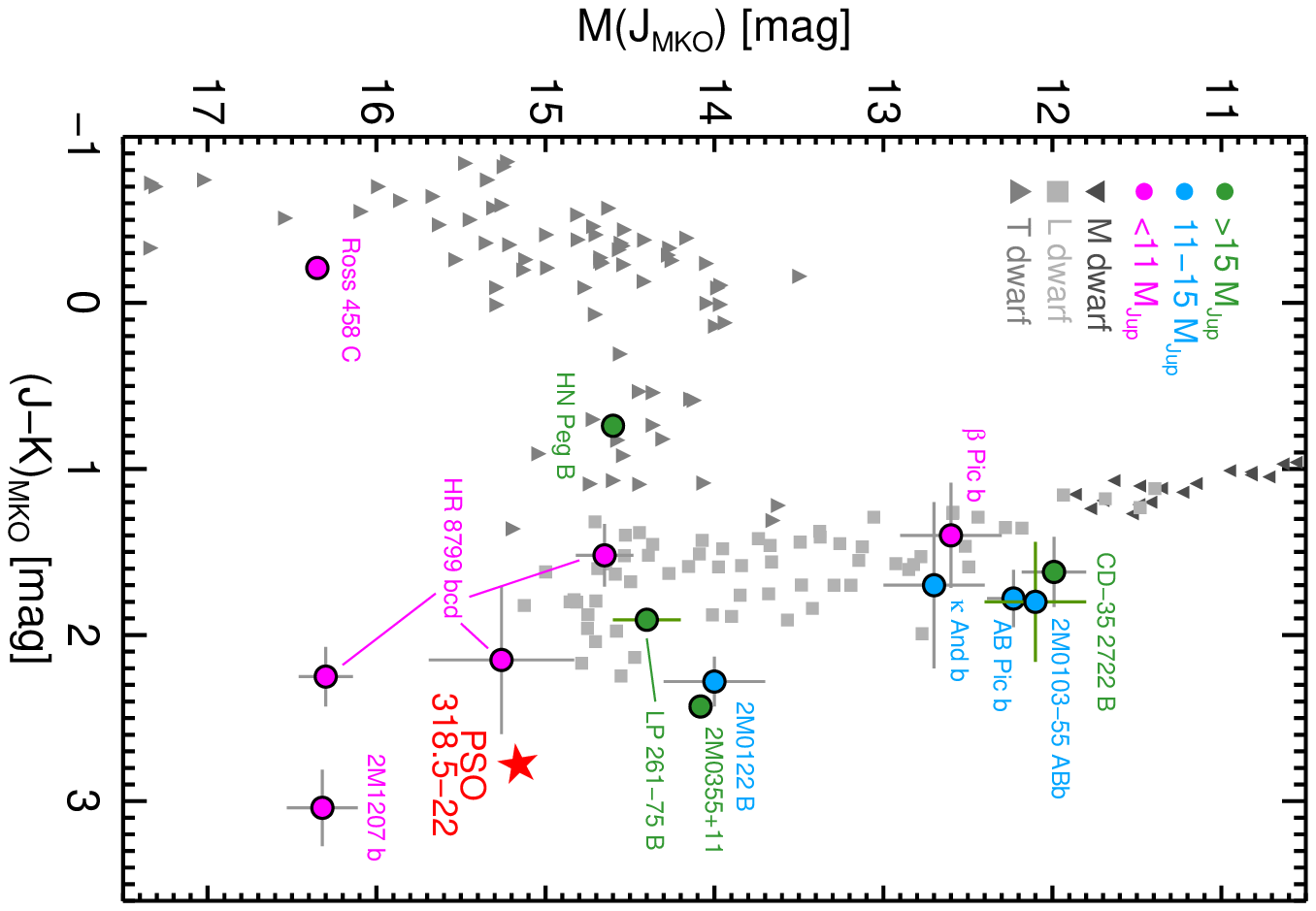}
  }
}
\vskip -7ex
\caption{{\bf Left:} Color-color diagrams using optical+IR
   ({\em top left}) and IR-only ({\em bottom left}) photometry. Field
  objects are plotted based on our PS1 photmetry ($\yps$) and the MKO
  photometry ($JHK$) compilation by \citet{2010ApJ...710.1627L}. The
  extreme colors of \psobject\ compared to the field population are
  evident. Also shown are the young planetary-mass object 2MASS
  1207$-$39b (Chauvin et al 2005) and the very red field L~dwarfs
  2MASS~J2148+40, 2MASS~J2244+20, 2MASS~J0355+11, and WISE~J0047+68.
  {\bf Right:} \psobject\ compared to known substellar objects based on
  the compilations of \citet{2012arXiv1201.2465D},
  \citet{2013ApJ...774...55B}, and references therein. Young substellar
  companions are highlighted, with the AB~Pic~b data from
  \citet{2013arXiv1309.1462B}. \psobject\ is very red and faint compared
  to field L~dwarfs, with magnitudes and colors comparable to the
  planets around HR~8799 and 2MASS~J1207$-$39. (Its measurements
  uncertainties are smaller than the symbol
  size.) \label{fig:colorcolor}}
\end{figure}

The colors of \psobject\ are extreme, with \mbox{$(J-K)_{MKO} =
  2.78$}~mag, \mbox{$(J-K)_{2MASS} = 2.84$}~mag, and
$(W1-W2)=0.76\pm0.04$~mag, all being the reddest among field L~dwarfs
(Figure~\ref{fig:colorcolor} and also see
\citealp{2012AJ....144...94G}). Such colors are thought to arise from an
unusually dusty atmosphere that results from a low surface gravity
(young age).
The position of \psobject\ on the near-IR color-magnitude diagram is
similarly extreme, being significantly fainter in $J$-band absolute
magnitude than field L~dwarfs (Figure~\ref{fig:colorcolor}). It
coincides with the colors and magnitudes of the directly imaged planets
around HR~8799 and 2MASS~J1207$-$39b.

We determine the near-IR spectral type of \psobject\ using the
\citet{allers13-young-spectra} system, which provides
gravity-insensitive types consistent with optical spectral types. For
late-L dwarfs, visual classification in the $J$~and $K$~bands and
index-based classifications with the H$_2$OD index are applicable. For
the GNIRS spectrum of \psobject, we visually assign a $J$-band type of
L9$\pm$1 and a $K$-band type of L6$\pm$1. The H$_2$OD index corresponds
to L6.0$\pm$0.8. The weighted mean of these three determinations leads
to a final type of L7$\pm$1. Spectral typing of our low-resolution SpeX
spectrum gives the same classification (Table~1).

\begin{figure}
  \vskip -0.3in
  \centering{\includegraphics[width=3.3in,angle=90]{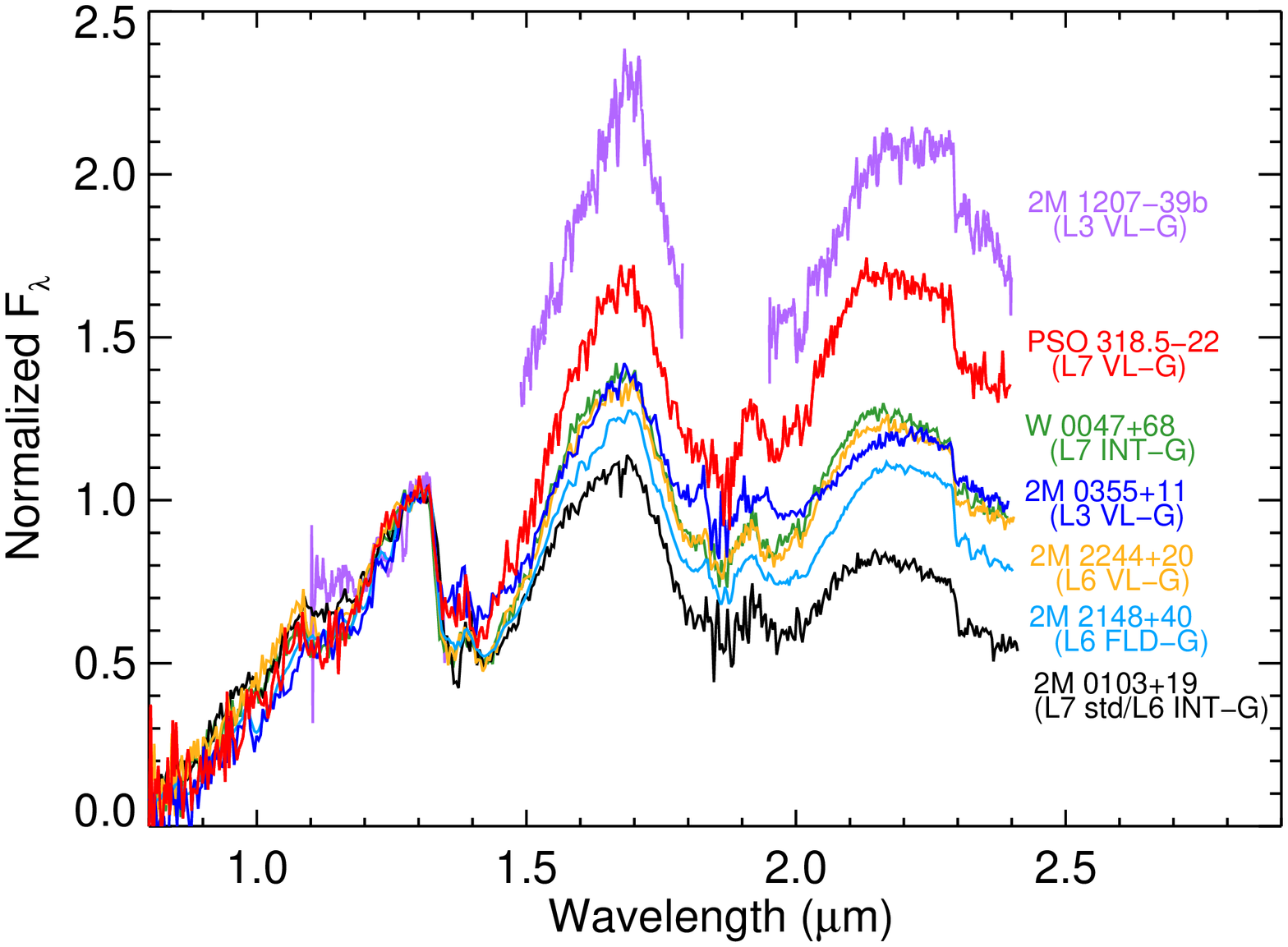}}
  %\centering{\includegraphics[width=3.3in,angle=90]{plot-compare.ps}}
  \centering{\hbox{
      \hskip 0.15in 
      \includegraphics[width=4.3in,angle=0]{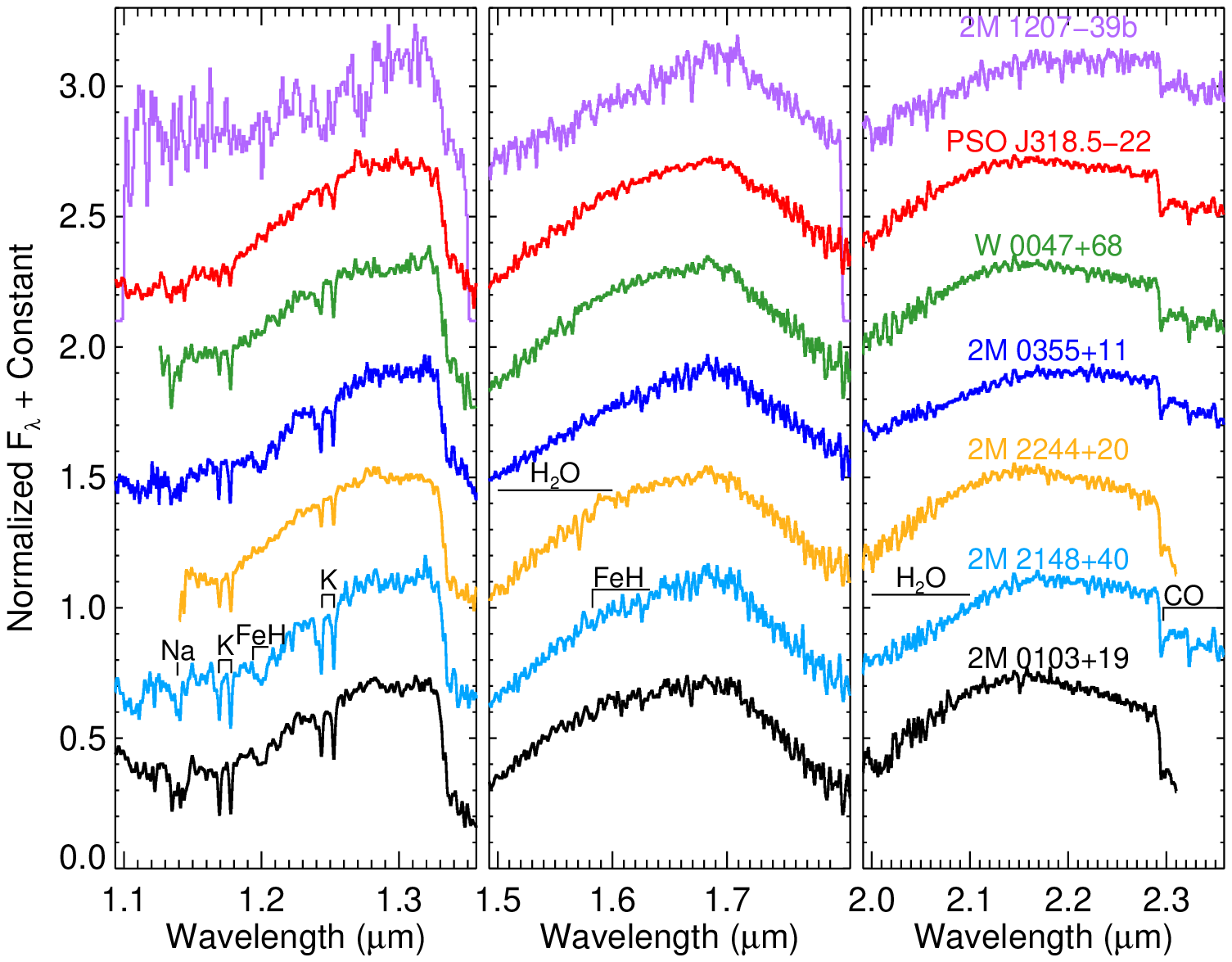}
    }
  }
  \caption{Spectrum of
    \psobject\ compared to the L7 near-IR standard 2MASS~J0103+19 of
    \citet{2010ApJS..190..100K}; the dusty field object 2MASS~J2148+40
    (\citealp{2008ApJ...686..528L}, which we classify as L6$\pm$1~{\sc
      fld-g}); the L6~{\sc vl-g} standard 2MASS~J2244+20 from
    \citet{allers13-young-spectra}; the very red L~dwarfs 2MASS~J0355+11
    \citep{2008AJ....136.1290R, 2013AJ....145....2F,
      allers13-young-spectra} and WISE~J0047+68
  (\citealp{2012AJ....144...94G}; Gizis \etal, in prep.); and the young
    planetary-mass object 2MASS~1207$-$39b \citep{2010A&A...517A..76P}.
    The labels use near-IR types and gravity classifications on the
    Allers \& Liu system, except for the \citet{2010ApJS..190..100K} L7
    standard (which Allers \& Liu classify as L6~{\sc int-g} in
    the near-IR).
    {\bf Top:} Low-resolution ($R\approx100$) spectra, normalized to the $J$-band peak 
    (1.26--1.31~\micron). 
    The 2MASS~J1207$-$39b spectrum has
    been lightly smoothed. Note that despite having similar colors,
    luminosities, and ages,  
    2MASS~J1207$-$39b and \psobject\ have very different $H$-band
    continuum shapes. 
    {\bf Bottom:}~Moderate-resolution spectra for the same young and/or
    dusty objects, all smoothed to $R=750$. The weaker $J$-band
    \ion{Na}{1} and \ion{K}{1} lines for \psobject\ compared to
    2MASS~J2244+20 and WISE~J0047+68 indicate a lower
    gravity. \label{fig:spectra}}
\end{figure}

\psobject\ shows a triangular $H$-band continuum, which is considered a
hallmark of youth \citep[e.g.][]{2001MNRAS.326..695L}. However,
\citet{allers13-young-spectra} caution that very red L-dwarfs having no
signatures of youth (low gravity) can display a triangular $H$-band
shape (\eg, 2MASS~J2148+40 in Figure \ref{fig:spectra}). At moderate
resolution, there are other indicators of youth for late-L dwarfs. Our
GNIRS spectrum displays a weak 1.20~$\mu$m FeH band as well as weak
\ion{Na}{1} (1.14~$\mu$m) and \ion{K}{1} (1.17 and 1.25~$\mu$m) lines,
which indicate a low gravity.
Using the gravity-sensitive indices of \citet{allers13-young-spectra},
we classify \psobject\ as {\sc vl-g}, which Allers \& Liu suggest
correspond to ages of $\sim$10--30~Myr based on the (small) sample of
young late-M/early-L dwarfs with good age constraints. Altogether,
\psobject\ visually appears most similar to the red L~dwarfs
WISE~J0047+68 \citep{2012AJ....144...94G} and 2MASS~J2244+20
\citep{2003ApJ...596..561M}, in accord with the similar spectral types
and gravity classifications for these three objects.

Allers \& Liu (2013) note that objects of the same age and spectral type
(temperature) can display different spectral signatures of youth,
based on 2~young L~dwarfs in the AB~Dor moving group. Our new
discovery affirms this idea. The spectra of 2MASS~J1207$-$39b and
\psobject\ are quite different (Figure \ref{fig:spectra}), despite their
similar colors and absolute magnitudes and the fact they may be coeval
(Section~\ref{sec:membership}). \psobject\ shows a negative continuum
slope from 2.12--2.28~$\mu$m, whereas 2MASS~J1207$-$39b has a positive
slope. The $H$-band continuum of \psobject\ displays a "shoulder" at
1.58~$\mu$m, whereas 2MASS~J1207$-$39b has a very peaked continuum.
Overall, the spectrum of 2MASS~J1207$-$39b appears most similar to that
of 2MASS~J0355+11 (both objects have near-IR types of L3 {\sc vl-g}),
despite their large differences in ages, colors, and absolute
magnitudes. Altogether, these comparisons hint that determining relative
ages and temperatures from NIR spectra may be unexpectedly complex.

To assess the physical parameters, we fit the low-resolution near-IR
spectra of \psobject\ and 2MASS~J0355+11 with the Ames/DUSTY model
atmospheres \citep{2001ApJ...556..357A} and the BT-Settl model
atmospheres \citep{2011ASPC..448...91A} with two different assumed solar
abundances (\citealp{2009ARA&A..47..481A} [AGSS] and
\citealp{2011SoPh..268..255C} [CIFIST]). Since both objects have
parallaxes, the scaling factors $(R^2/d^2)$ from the fits provide an
estimate of the objects' radii.
Using \chisq\ minimization, all three sets of models indicates that
\psobject\ ($\Teff_{,atm}=1400-1600~K$) is 100--200~K cooler than
2MASS~J0355+11 ($\Teff_{,atm}=1600-1700~K$). BT-Settl/AGSS models give
the best fits,
with the resulting temperatures being comparable to previous fitting of
field L~dwarfs but with both young objects having lower surface
gravities (\logg=4.0--4.5~dex) than field objects
(\citealp{2007arXiv0711.0801C, stephens09-irac, 2009A&A...503..639T},
though the number of fitted field objects is small). Most strikingly,
the fitted radii are implausibly small ($\approx$0.8~\Rjup), indicating
discord between the data and models even though the quality of fit seems
adequate by eye. A similar radius mismatch occurs in model atmosphere
fitting of young dusty planets (\eg, \citealp{bowler10-hr8799b,
  2011ApJ...733...65B}; \cf, \citealp{2012ApJ...754..135M}).

\subsection{Group Membership \label{sec:membership}}

\begin{figure}
\hskip -0.6in
\includegraphics[width=5.2in,angle=90]{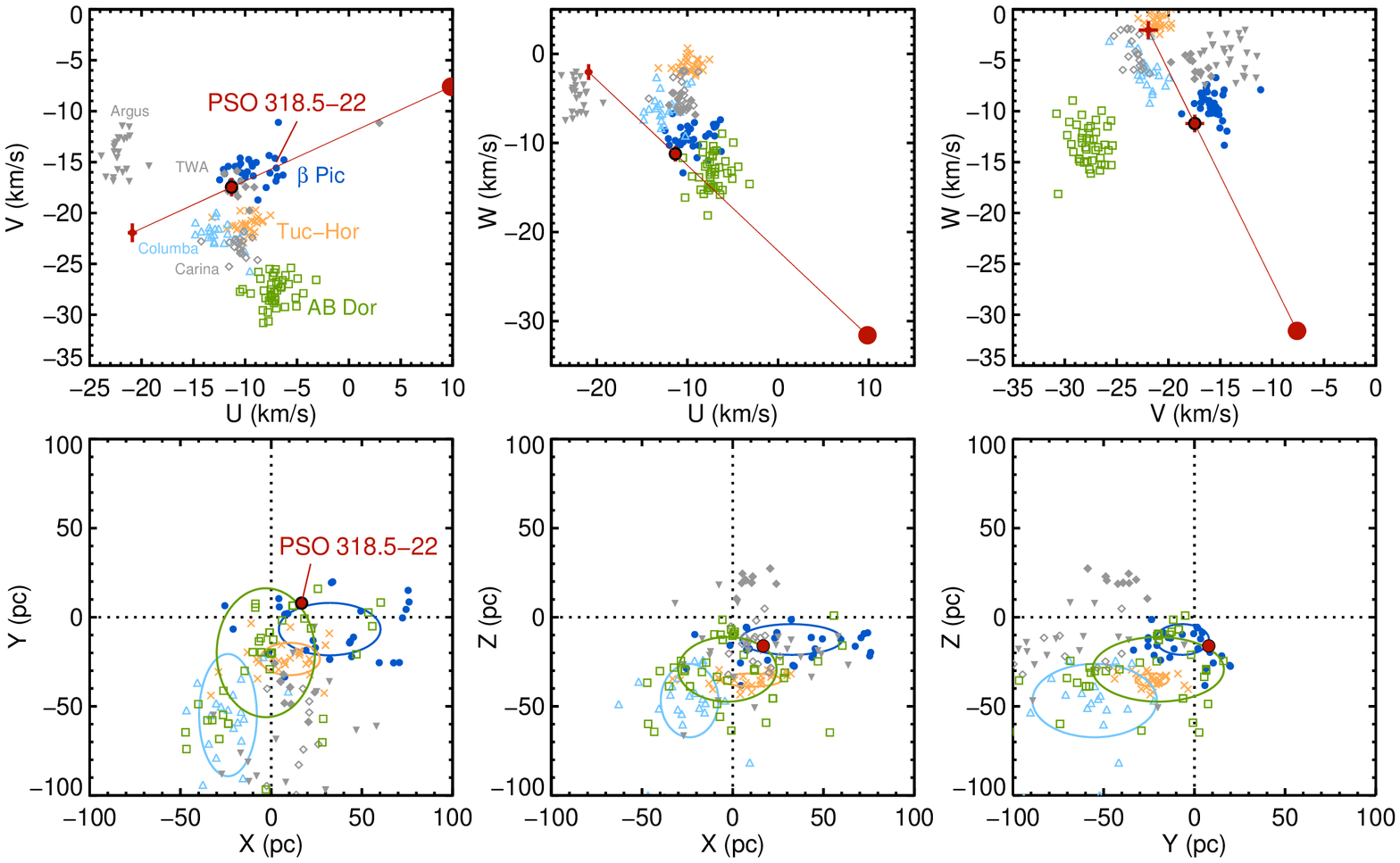}
\vskip -0.5in
\caption{Kinematic and spatial position of \psobject\ (red circle)
  compared to known YMGs from \citet{2008hsf2.book..757T}. For the 
  $UVW$ plots, we adopt an RV of \hbox{[$-$20, +25]~\kms}
  (Section~\ref{sec:membership}), with the larger red circles being more
  positive velocities and the red line being the full RV range. 
  Uncertainties arising from errors in the distance and proper motion
  (but not the RV) are show at the 3 red circles. The members of the
  nearest YMGs to \psobject\ are shown in color and the others in
  grey, with the ellipses representing the RMS of the known members.
  Consideration of both the $XYZ$ and $UVW$ data indicate the \bPic\
  group is the only possibility. The middle (black-outlined) circle
  shows the $UVW$ position of \psobject\ for an RV of
  $-$6~\kms. \label{fig:uvwxyz}}
\end{figure}

There is no radial velocity (RV) for \psobject, but we can place
  modest constraints on its space motion. Published RVs
  \citep{2010ApJ...723..684B, 2010A&A...512A..37S, 2010ApJS..186...63R,
    2012ApJ...758...56S} find that low-gravity late-M and L~dwarfs
  reside in the range of [$-$20, +25]~\kms, with the smaller RV range
  compared to field objects being expected given the young ages.
Figure~\ref{fig:uvwxyz} plots the spatial ($XYZ$) and kinematic ($UVW$)
location of \psobject\ compared to the young moving groups (YMGs) from
\citet{2008hsf2.book..757T}.\footnote{$U$ and $X$ are positive toward
  the Galactic Center, $V$ and $Y$ are positive toward the direction of
  galactic rotation, and $W$ and $Z$ are positive toward the North
  Galactic Pole.} Its $XYZ$ position is coincident with the \bPic,
Tuc-Hor, and AB~Dor moving groups. However, the velocity information
makes the Tuc-Hor and AB~Dor groups implausible, given the large offset
from \psobject. Overall, the match with the \bPic\ members is excellent,
\ie, the velocity offset between the group and \psobject\ would be only
3~\kms\ if \psobject\ has an RV of $-$6~\kms.

We also computed the kinematic distance ($d_{kin}$) and angle $\theta$
from
\citet{Schlieder2010}.
$d_{kin}$ is an object's distance assuming that its tangential velocity
(as derived from its observed proper motion) is the same as the mean
tangential velocity of a YMG. Including the uncertainties in the \bPic\
group's $UVW$, \psobject\ has $d_{kin}=20.8^{+1.1}_{-1.0}$~pc,
in good agreement with its parallactic distance.
$\theta$ gives the angle between the object's proper motion and that of
the YMG at the object's sky position. \psobject\ has
$\theta=8^{+4}_{-5}$, whereas the \bPic\ members from
\citet{2008hsf2.book..757T} mostly have $\theta<15$~degs. In other
words, for \psobject\ the amplitude and direction of its tangential
velocity given its sky location agrees well with the space motion of
known \bPic\ members.

Finally, we computed the YMG membership probability using the online
calculator of \citet{2013ApJ...762...88M}.
Given the position and parallax of \psobject, their Bayesian method
reports a membership probability of 99.9\% in the \bPic\ YMG and 0.01\%
in the field.
(Removing the parallax as an input, their method gives 99.6\% and a
predicted distance of 22.5$\pm$1.6~pc, in good agreement with our
parallactic distance.)
The absolute value of this method's probabilities is imperfect, since
their input model for the solar neighborhood does not account for the
relative numbers of objects in different YMGs and the field. However,
the high probability for \psobject\ is comparable to the values computed
for most of the known \bPic\ members (and exceeds the rest of them). The
Malo \etal\ method also does not consider the fact that \psobject\ is
spectroscopically young, which would boost the probability that it is a
YMG member rather than a field object (since most field objects have old
ages).

\subsection{Physical Properties \label{sec:properties}}

\psobject, 2MASS~J0355+11, and 2MASS~J1207$-$39b are all moving
group members and thus play an important role as ``age benchmarks''
\citep[e.g.][]{2006MNRAS.368.1281P, liu08-2m1534orbit} for very red
L~dwarfs. With a measured \Lbol\ and an age estimate, we use the
\citet{2008ApJ...689.1327S} evolutionary models with $f_{sed}=2$ to
compute the remaining physical properties. (Results from the other
commonly available evolutionary models are similar, comparable to the
uncertainties.)

To derive \Lbol, we combined our 0.9--2.4~\micron\ low-resolution
spectra flux-calibrated using the $JHK$ photometry, the \WISE\ $W1$
  and $W2$ photometry, and the best-fitting atmospheric model
  (Section~\ref{sec:spectra}) for the flux at bluer and redder
  wavelengths. We integrated the fluxes and used the parallaxes to
determine \Lbol, with the uncertainties in the input data (spectra,
photometry, and parallax) handled in a Monte Carlo fashion. We find
$\log(\Lbol/\Lsun)=-4.42\pm0.06$~dex and $-4.23\pm0.11$~dex for
\psobject\
and 2MASS~J0355+11, respectively, very similar to field mid/late-L
dwarfs \citep{2004AJ....127.3516G, 2006ApJ...648..614C}. Young, very red
L~dwarfs are sometimes (and confusingly) described as
  ``underluminous'' compared to field objects, but we find that the
  luminosities of young and old field objects are
  comparable.\footnote{Note that 2MASS~J1207$-$39b can be
    described as ``underluminous,'' as its luminosity
    \citep[$\log(\Lbol/\Lsun)=-4.73\pm0.12$~dex;][]{2011ApJ...735L..39B}
    is somewhat low relative to field L~dwarfs.} Rather, these objects
have fainter absolute magnitudes in the bluer near-IR bands (\eg,
$J$~band) but comparable \Lbol's. They are therefore better described as
being ``very red.''

\begin{figure}
\vskip -3in
\hskip -2in
\includegraphics[width=8in,angle=90]{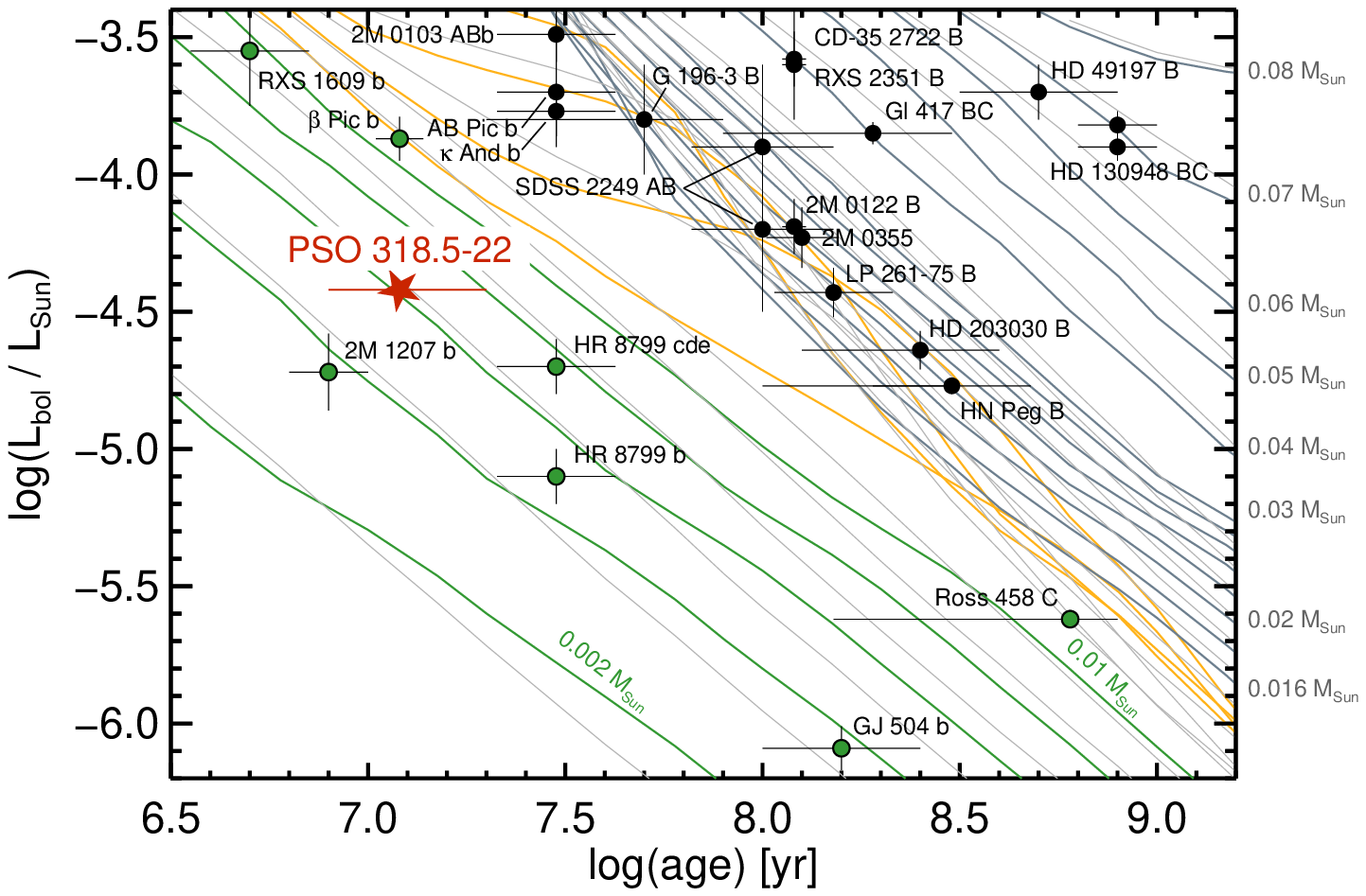}
\vskip -1.3in
\caption{\psobject\ and known young substellar objects from
  \citet{2013ApJ...774...11K} and the compilation of
  \citet{2013ApJ...774...55B}. The \citet{2008ApJ...689.1327S} cloudless
  evolutionary models are plotted in color, with 0.002--0.010~\Msun\
  tracks in green (spaced by 0.002~\Msun), 0.011-0.014~\Msun\ tracks in
  yellow (where D-burning may complicate the interpretation and spaced
  by 0.001~\Msun), and $\ge$0.016~\Msun\ in grey (spaced by 0.002~\Msun\
  up to 0.030~\Msun\ and then by 0.010~\Msun). The thin light grey lines
  show the cloudy evolutionary models with $f_{sed}=2$. \label{fig:evolmodels}}
\end{figure}

For \psobject, using an age of 12$^{+8}_{-4}$~Myr (distributed
uniformly) from \citet{2001ApJ...562L..87Z}, we find a mass of
$6.5^{+1.3}_{-1.0}$~\Mjup, temperature of 1160$^{+30}_{-40}$~K, and
\logg\ of 3.86$^{+0.10}_{-0.08}$~dex (Figure~\ref{fig:evolmodels}).
Adopting a larger age range of 10--100~Myr (distributed uniformly), the
mass becomes 12$\pm$3~\Mjup, with \Teff\ and \logg\ increasing only
slightly (Table~1). Such cool temperatures correspond to a field
$\approx$T5.5~dwarf \citep{2004AJ....127.3516G} and yet \psobject\ shows
no sign of \meth\ in its spectrum, similar to the situation for the
planets around HR~8799 and 2MASS~J1207$-$39.
\citep[e.g.][]{marois08-hr8799bcd, bowler10-hr8799b,
  2010A&A...517A..76P, 2011ApJ...733...65B}. The evolutionary-model
radii are almost twice as large as those from our model atmosphere fits,
as has previously been noted in analysis of directly imaged planets
(Section~\ref{sec:spectra}).

Our same calculations for 2MASS~J0355+11 give a mass of
  $24^{+3}_{-6}$~\Mjup, temperature of 1420$^{+80}_{-130}$~K, and \logg\
  of $4.58^{+0.07}_{-0.17}$~dex, assuming an age of 125$\pm$25~Myr
  \citep{2013ApJ...766....6B} based on its membership in the AB~Dor
  moving group \citep{2013AJ....145....2F, 2013AN....334...85L}. This
  object is more massive and hotter than \psobject, as expected given
  its brighter absolute magnitudes and older age.
  \citet{2013AJ....145....2F} describe 2MASS~J0355+11 as having a strong
  resemblance to a giant exoplanet. While it does have similar $H$~and
  $K$-band spectra to 2MASS~J1207$-$39b (but not in $J$~band nor the
  overall SED shape; Figure~\ref{fig:spectra}), 2MASS~J0355+11 is
  $\approx$2~mag brighter in $J$~band (Figure~\ref{fig:colorcolor}) and
  $\approx$5$\times$ more massive than 2MASS~J1207$-$39b. Likewise, the
  spectrum of HR~8799b has been shown to resemble some field dwarfs
  (\eg, 2MASS~J2148+40 and 2MASS~J2244+20 by \citealp{bowler10-hr8799b}
  and 2MASS~J2139+02 by \citealp{2011ApJ...733...65B}), yet HR~8799b is
  much fainter and lower mass. This strengthens our finding that {\em
    near-IR spectral morphology can not reliably determine the
    underlying physical properties of very red young objects}.

The temperatures for these two objects are substantially lower then
those measured in the same fashion (using \Lbol, age, and evolutionary
models) for field L~dwarfs of comparable spectral types.
\citet{2004AJ....127.3516G} find \Teff\ of 1700--1950~K for spectral
types L3--L5 (corresponding to the near-IR and optical types for
2MASS~J0355+11) and 1500~K for L7 (the near-IR type for \psobject).
Thus, the temperatures of these young, very red L~dwarfs are
$\approx$400~K cooler than comparable field objects.
\citet{2013ApJ...774...55B} show that young L-type companions tend to
have cooler temperatures than (old) field objects of the same type. Our
results show that this offset also occurs for young free-floating
objects.

%----------------------------------------------------------------------%
\section{Discussion \label{sec:discussion}}

\psobject\ shares a strong physical similarity to the young dusty
planets HR~8799bcd and 2MASS~J1207$-$39b, as seen in its colors,
absolute magnitudes, spectrum, luminosity, and mass. Most notably,
  it is the first field L~dwarf with near-IR absolute magnitudes as
  faint as the HR~8799 and 2MASS~J1207$-$39 planets, {\em demonstrating
    that the very red, faint region of the near-IR color-magnitude
    diagram is not exclusive to young exoplanets.} Its probable
membership in the \bPic\ moving group makes it a new substellar
benchmark at young ages and planetary masses. We find very red,
low-gravity L~dwarfs have $\approx$400~K cooler temperatures relative to
field objects of comparable spectral type, yet have similar
luminosities. Comparing very red L~dwarf spectra to each other and to
directly imaged planets highlights the challenges of diagnosing physical
properties from near-IR spectra.

\psobject\ is among the lowest-mass free-floating objects identified in
the solar neighborhood. \citet{2013arXiv1309.1422D} have determined
parallaxes and luminosities of field Y~dwarfs and thereby estimate
masses of 7--20~\Mjup, assuming ages of 1--5~Gyr. The (likely Y~dwarf)
companion WD~0806$-$661B has a precise age from its white dwarf primary,
leading to a mass of 6--10~\Mjup\ \citep{2012ApJ...744..135L,
  2013arXiv1309.1422D}. \citet{2012arXiv1210.0305D} have identified a
candidate young late-T dwarf member of the AB~Dor moving group with an
estimated mass of 4--7~\Mjup, though a parallax and RV are needed to
better determine its properties and membership (and thus age and mass).

\psobject\ was not discovered by previous large-area searches for
L~dwarfs using 2MASS \citep{2008AJ....136.1290R, 
  2003AJ....126.2421C}.
\psobject\ has only tenuous detections in the 2MASS $J$~and $H$~bands,
and the resulting colors lie outside the selection criteria of
\citet{2003AJ....126.2421C}. Additionally, the object is too far south
to be detected by SDSS,
and it is not in the areas surveyed so far by UKIDSS
or VISTA.
However, both the PS1 and VISTA datasets are growing relative to what
has been searched here, both in depth and area, and upcoming surveys
(\eg, LSST) will have even greater reach. Thus, wide-field surveys
mapping the whole sky offer a promising avenue for understanding
exoplanets directly imaged in the tiny areas around the brightest stars.

%----------------------------------------------------------------------%
\acknowledgments

The Pan-STARRS1 surveys have been made possible by the Institute for
Astronomy, the University of Hawaii, the Pan-STARRS Project Office, the
institutions of the Pan-STARRS1 Science Consortium ({\tt
  http://www.ps1sc.org}), NSF, and NASA.
We thank Brendan Bowler for assistance with the figures and Michael
  Cushing for providing a pre-release update of SpeXtool. Our research
has employed the \WISE\ and 2MASS data products; NASA's Astrophysical
Data System;
and the Spex Prism Spectral Libraries maintained by Adam Burgasser.
This research was supported by NSF grants AST09-09222 (awarded to MCL)
and AST-0709460 (awarded to EAM) as well as AFRL Cooperative Agreement
FA9451-06-2-0338
Finally, the authors wish to recognize and acknowledge the very
significant cultural role and reverence that the summit of Mauna Kea has
always had within the indigenous Hawaiian community. We are most
fortunate to have the opportunity to conduct observations from this
mountain.

{\it Facilities:} \facility{IRTF (SpeX), CFHT (WIRCAM)}

\clearpage
%\bibliography{/users/mliu/tex/bibtex/mliu}
%\bibliographystyle{apj}

%======================================================================%
\clearpage
%% use the figure environment and \plotone or \plottwo to include 

%==========================================================================================%
\clearpage

\begin{deluxetable}{lccccccc}
\tablecaption{Measurements of \psobjectfull\ \label{table}}
\tabletypesize{\small}
%\rotate
\tablewidth{0pt}
%\tablecolumns{2}
\tablehead{
  \colhead{Property} &
  \colhead{Measurement} 
}

\startdata
%----------------------------------------%
\cutinhead{Nomenclature}
%----------------------------------------%
2MASS                      &  2MASS~J21140802$-$2251358 \\
\ps\                       &  PSO~J318.5338$-$22.8603   \\   
\WISE\                     &  WISE~J211408.13$-$225137.3  \\
%----------------------------------------%
\cutinhead{Astrometry (Equinox J2000)}
%----------------------------------------%
2MASS RA, Dec (ep 1999.34)   &  318.53344, $-$22.85996  \\   
\ps\   RA, Dec  (ep 2010.0)   &  318.53380, $-$22.86032  \\  
\WISE\ RA, Dec  (ep 2010.34)  &  318.53388, $-$22.86037  \\  

Proper motion $\mu_\alpha, \mu_\delta$ (mas/yr)  & 137.3 $\pm$ 1.3, $-$138.7 $\pm$ 1.4 \\
Proper motion amplitude $\mu$ (\arcsec/yr)     & 195.0 $\pm$ 1.3 \\
Proper motion PA (\degs)                       & 135.3 $\pm$ 0.4 \\
Parallax $\pi$ (mas)                           & 40.7 $\pm$ 2.4  \\ 
Distance $d$ (pc)                              & 24.6 $\pm$ 1.4  \\
$v_{tan}$ (km/s)                                & 22.7 $\pm$ 1.3  \\
%----------------------------------------%
\cutinhead{Photometry} 
%----------------------------------------%
PS1 \gps\ (AB mag)      & $>$23.6 (3$\sigma$)\tablenotemark{a} \\ 
PS1 \rps\ (AB mag)      & $>$23.1 (3$\sigma$)\tablenotemark{a} \\ 
PS1 \ips\ (AB mag)      & $>$22.9 (3$\sigma$)\tablenotemark{a} \\ 
PS1 \zps\ (AB mag)      & 20.80 $\pm$ 0.09\tablenotemark{a} \\   
PS1 \yps\ (AB mag)      & 19.51 $\pm$ 0.07\tablenotemark{a} \\   
2MASS $J$ (mag)         & 16.71 $\pm$ 0.20 \\
2MASS $H$ (mag)         & 15.72 $\pm$ 0.17 \\
2MASS $\Ks$ (mag)       & 14.74 $\pm$ 0.12 \\
MKO $Y$ (mag)           & 18.81 $\pm$ 0.10   \\
MKO $J$ (mag)           & 17.15 $\pm$ 0.04   \\  
MKO $H$ (mag)           & 15.68 $\pm$ 0.02   \\
MKO $K$ (mag)           & 14.41 $\pm$ 0.02   \\  
WISE $W1$ (mag)         & 13.22 $\pm$ 0.03  \\
WISE $W2$ (mag)         & 12.46 $\pm$ 0.03  \\
WISE $W3$ (mag)         & 11.8\phn $\pm$ 0.4\phn  \\
WISE $W4$ (mag)         & $>$8.6 (2$\sigma$) \\
%----------------------------------------------------------------------%
\cutinhead{Synthetic photometry\tablenotemark{b}}
%----------------------------------------------------------------------%
2MASS $J-H$   (mag) &  1.678 $\pm$ 0.007 (0.06)  \\
2MASS $H-\Ks$ (mag) &  1.159 $\pm$ 0.004 (0.03)  \\
2MASS $J-\Ks$ (mag) &  2.837 $\pm$ 0.006 (0.07)  \\
MKO $Y-J$    (mag)  &  1.37\phn $\pm$ 0.02\phn (0.05)  \\
MKO $J-H$    (mag)  &  1.495 $\pm$ 0.007 (0.06)  \\
MKO $H-K$    (mag)  &  1.279 $\pm$ 0.004 (0.03)  \\
MKO $J-K$    (mag)  &  2.775 $\pm$ 0.007 (0.07)  \\
$(J_{2MASS} - J_{MKO})$ (mag)   & \phs0.111 $\pm$ 0.002  \\
$(H_{2MASS} - H_{MKO})$ (mag)   & $-$0.072 $\pm$ 0.001  \\
$(\Ks_{,2MASS} - K_{MKO})$ (mag) & \phs0.049 $\pm$ 0.001  \\
log(\Lbol/\Lsun) (dex)     &  $-$4.42 $\pm$ 0.06 \\
%----------------------------------------------------------------------%
\cutinhead{Spectral classification\tablenotemark{c}}
%----------------------------------------------------------------------%
$J$-band type (SpeX, GNIRS)  &  L8   $\pm$ 1,     L9 $\pm$ 1    \\
$H$-band type (SpeX, GNIRS)  &  L6   $\pm$ 1,     L6 $\pm$ 1    \\
H$_2$OD type (SpeX, GNIRS)   &  L6.0 $\pm$ 0.9, L6.0 $\pm$ 0.8  \\
Gravity score (SpeX, GNIRS)  & XXX2, 2X21 \\
Final near-IR spectral type  & L7 $\pm$ 1 \\
Near-IR gravity class    & {\sc vl-g} \\      
%
%----------------------------------------------------------------------%
\cutinhead{Physical Properties (age = 12$_{-4}^{+8}$~Myr)}
%----------------------------------------------------------------------%
Mass (\Mjup)             &  6.5 ($-$1.0, +1.3)     \\   
\Teff$_{,evol}$ (K)       &  1160 ($-$40, +30)      \\   
$\log(g_{evol})$ (cgs)    &  3.86 ($-$0.08, +0.10)  \\
Radius (\Rjup)           &  1.53 ($-$0.03, +0.02)  \\
%
%----------------------------------------------------------------------%
\cutinhead{Physical Properties (age = 10--100~Myr)}
%----------------------------------------------------------------------%
Mass (\Mjup)             &  12 $\pm$ 3        \\      
\Teff$_{,evol}$ (K)       &  1210 ($-$50, +40) \\   
$\log(g_{,evol})$ (cgs)   &  4.21 ($-$0.16, +0.11)  \\
Radius (\Rjup)           &  1.40 ($-$0.04, +0.06)  \\
\enddata

\tablenotetext{a}{Average of multi-epoch photometry. 
  The optical non-detections are consistent with the colors of late-L
  dwarfs (\hbox{$\ips-\zps\approx2.0-2.5$~mag};
  \citealp{2013arXiv1309.0503B}).}

\tablenotetext{b}{Colors were synthesized from our SpeX spectrum, with
  formal errors derived from the spectrum's measurement errors.
  The values in parentheses are the estimated systematic errors,
  calculated from comparing the measured $JHK$ colors for
  $\sim$100~objects to the colors synthesized from the SpeX Prism
  Library. This additional uncertainty is needed to reconcile the two
  sets of results such that $P(\chi^2) \approx 0.5$. No systematic
  errors ($\lesssim$0.02~mag) were measurable for the 2MASS-to-MKO
  conversion within a given bandpass \citep{2012arXiv1201.2465D}.}

\tablenotetext{c}{Based on the classification system of
  \citet{allers13-young-spectra}.
    }

\end{deluxetable}
%==========================================================================================%

\end{document}